\documentclass[12pt]{article}
\usepackage{graphicx}
\usepackage{amssymb}
\usepackage{amsmath}
\usepackage{amsfonts}
\usepackage{mathtools}
\usepackage{multirow}
\usepackage{url}
\usepackage{alltt}
\usepackage{float}

\newtheorem{definition}{Definition}

\begin{document}

\title{\texttt{CGAlgebra}: a \emph{Mathematica} package for conformal
    geometric algebra. v.2.0}

\author{\textbf{E. Alejandra Ort\'{\i}z-Dur\'an} and \textbf{Jos\'e L. Arag\'on}}
\date{Centro de F\'{\i}sica Aplicada y Tecnolog\'{\i}a
    Avanzada, \\
    Universidad Nacional Aut\'{o}noma de M\'{e}xico, \\
    Apartado Postal 1-1010, Quer\'{e}taro 76000, Mexico\\
    \texttt{jlaragon@unam.mx}}
  
\maketitle
  

\begin{abstract} 
  A tutorial of the \textit{Mathematica}$^{\textregistered}$ package
  \texttt{CGAlgebra}, for conformal geometric algebra calculations is
  presented. Using rule-based programming, the 5-dimensional conformal
  geometric algebra is implemented and the defined functions simplify
  the calculations of geometric, outer and inner products, as well as
  many other calculations related with geometric
  transformations. \texttt{CGAlgebra} is available from
  \url{https://github.com/jlaragonvera/Geometric-Algebra}
\end{abstract}

\section{Introduction} 

In the 5D conformal geometric algebra geometric objects such as lines,
planes, circles and spheres in 3D are represented in a simple way by
algebraic identities. It is an extension of the 4D projective
geometric algebra and was proposed by D. Hestenes \cite{Hestenes0,Li}
as a powerful framework for computational Euclidean
geometry. Extensive applications of the conformal geometric algebra to
computer graphics, computer vision and robotics have been reported. As
general references we recommend the books
\cite{Dorst,Perwass,Hildenbrand,Kanatani}.

\begin{definition}[The conformal space] The five-dimensional space
  spanned by $\left\{ e_0, e_1, e_2, e_3, e_\infty \right\}$ with the
  inner product
\begin{equation}
 \label{eqn:confip}
 \langle {\mathbf x}, {\mathbf y} \rangle = x_1 y_1 + x_2 y_2 + x_3
 y_3 - x_0 y_\infty - x_\infty y_0 ,
\end{equation}
for
${\mathbf x} = x_0 e_0 + x_1 e_1 + x_2 e_2 + x_3 e_3 + x_\infty
e_\infty$, and
${\mathbf y} = y_0 e_0 + y_1 e_1 + y_2 e_2 + y_3 e_3 + y_\infty
e_\infty$, vectors in this space, is called the \emph{conformal
  space}.
\end{definition}

From \eqref{eqn:confip} we see that
$\| e_0 \|^2 = \langle e_0, e_0 \rangle = 0$ and similarly
$\| e_\infty \|^2 =0$, so $e_0$ and $e_\infty$ are \emph{null
  vectors}.

The inner product \eqref{eqn:confip} has signature $(4,1)$ and the
conformal space is the Minkowsky space $\mathbb{R}^{4,1}$.

For the purpose of the \textit{Mathematica} package developed here, a
definition of the conformal geometric algebra based on generators and
relations is more suitable.

\begin{definition}[$\mathbb{G}^{4,1}$]
  The conformal geometric algebra $\mathbb{G}^{4,1}$ over
  $\mathbb{R}^{4,1}$, equipped with the inner product
  \eqref{eqn:confip}, is an algebra generated by the identity 1 and
  the symbols $\{ e_0, e_1, e_2, e_3, e_\infty \}$, by an associative
  multiplication operation called \emph{geometric product}, and the
  following relations:
\begin{subequations}
  \label{eqn:relations41}
  \begin{align}
    e_i e_j +  e_j e_i &= 0 ,\\
    e_i e_0 +  e_0 e_i &= 0 , \\
    e_i e_\infty +  e_\infty e_i &= 0, \\
    e_i ^2 &= 1 ,\\
    e_0 e_\infty + e_\infty e_0 &= -2 , \\
    e_0 ^2 = e_\infty ^2 &= 0 ,
  \end{align}
\end{subequations}
for $i,j = 1,2,3$.
\end{definition}

Another way to build the conformal geometric algebra
$\mathbb{G}^{4,1}$ is to start with the Clifford algebra
$\mathcal{C}^{4,1}$ over $\mathbb{R}^{4,1}$, which is the
five-dimensional space generated by $\{ e_1, e_2, e_3, e_4, e_5 \}$
with the inner product
\begin{equation}
 \langle {\mathbf x}, {\mathbf y} \rangle = x_1 y_1 + x_2 y_2 + x_3
 y_3 + x_4 y_4 - x_5 y_5 ,
\end{equation}
for ${\mathbf x} = x_1 e_1 + x_2 e_2 + x_3 e_3 + x_4 e_4 + x_5 e_5$,
and ${\mathbf y} = y_1 e_1 + y_2 e_2 + y_3 e_3 + y_4 e_5 + y_5 e_5$,
vectors in $\mathcal{C}^{4,1}$. In this space we now define the null vectors
\begin{equation}
  \label{eqn:e0einf}
e_0 = \frac{e_4 + e_5}{2}, \;\;\; \textrm{and} \;\;\;  e_\infty = e_5 - e_4 .
\end{equation}

With this definition, the conformal geometric algebra
$\mathbb{G}^{4,1}$ is the generated by
$\{ e_0, e_1, e_2, e_3, e_\infty \}$.

The inverse of the transformation \eqref{eqn:e0einf} is
\begin{equation}
e_4 = e_0 - \frac{e_\infty}{2}, \;\;\; \textrm{and} \;\;\;  e_5 = e_0 + \frac{e_\infty}{2}.
\end{equation}

Both approaches are used in the \textit{Mathematica} package.

Examples of the representation and transformations of the basic
geometric objects of this conformal space, using \texttt{CGAlgebra}
will be given in what follows.

\section{Getting started with \texttt{CGAlgebra}}
\label{sec:package}

\texttt{GCAlgebra} is freely available and can be downloaded from

\vspace{0.25cm}
\url{https://github.com/jlaragonvera/Geometric-Algebra}
\vspace{0.25cm}

In a \emph{Mathematica} session the package is loaded with the
command
\begin{alltt}
<<"DIR/CGAlgebra.m"
\end{alltt}
where \texttt{DIR} is the full path of the directory where the package
is located.

In \texttt{CGAlgebra}, a basis element of the conformal space is
denoted by \texttt{e[i]}. With this notation, $\mathbb{G}^{4,1}$ is
generated by \texttt{\{e[0],e[1],e[2],e[3],e[$\infty$]\}}. The
geometric product between basis elements $e_i e_j \cdots e_k$ will be
denoted as \texttt{e[i,j,....,k]}. To warm up we can test some
of the relations \eqref{eqn:relations41}:
\begin{alltt}
In[1]:= e[2,1]
Out[1]:= -e[1,2]
In[2]:= e[\(\infty\),0]
Out[2]:= -2-e[0,\(\infty\)]
In[3]:= e[\(\infty\),\(\infty\)]
Out[3]:= 0
In[3]:= e[0,0]
Out[3]:= 0
\end{alltt}
The geometric product between an arbitrary number basis elements is,
for example:
\begin{alltt}
In[4]:= e[1,\(\infty\),2,0]
Out[4]:= 2e[1,2]+e[0,1,2,\(\infty\)]
\end{alltt}

\begin{table}[h!]
  \caption{\label{tbl:t1}Basic functions of \texttt{CGAlgebra}}
\begin{center}
\begin{tabular}{|l|l|}
\hline 
\textbf{Expression} & \textbf{Output}\tabularnewline
\hline 
\hline 
\multirow{2}{*}{\texttt{GeometricProduct[A,B,C,...]}} &The
                                          geometric product of the\\
 ~ &   multivectors $A$, $B$,  $C$, $\ldots$ \\
\hline 
\multirow{2}{*}{\texttt{OuterProduct[A,B,C,...]}} & The outer
                                                    (Grasmann) product \\
~ &    of the multivectors $A$, $B$, $C$, $\ldots$ \\ 
\hline 
\multirow{2}{*}{\texttt{InnerProduct[A,B]}} & The inner
                                              product (left contraction)\\
~ & of the multivectors $A$ and $B$. \\
\hline 
\texttt{Grade[A,k]} & The $k$-vector part of the multivector $A$.\\
\hline 
\end{tabular}
\end{center}
\end{table}

To operate with arbitrary multivectors, the geometric product between
basis elements is extended by linearity to all $\mathbb{G}^{4,1}$. The
basic functions defined in \texttt{CGAlgebra} are listed in Table \ref{tbl:t1}.

As an example consider the geometric product between the multivectors
$A = e_1e_2e_3 + a \; e_\infty e_3 e_2$, $B = a \; e_2$, $C=3$ and
$D = 4 + e_1 e_3$. It is computed as
\begin{alltt}
In[5]:= GeometricProduct[e[1,2,3]+a e[\(\infty\),3,2], a e[2],3,4+e[1,3]]
Out[5]:= 3a - 12a e[1,3] + 3a^2 e[1,\(\infty\)] - 12a^2 e[3,\(\infty\)] 
\end{alltt}

The expressions \texttt{e[]}, \texttt{GeometricProduct[]} and
\texttt{Grade[]} constitute the basis of the \texttt{CGAlgebra}
package. Extra defined functions and functionality will be presented
through examples concerning the representation and transformation of
geometrical objects in the conformal space.

It is important to mention that all the results are given in terms of
geometric products of the basis vectors
\texttt{\{e[0],e[1],e[2],e[3],e[$\infty$]\}}. Thus, for instance, the
pseudoscalar $I_5 = e_0 \wedge e_1 \wedge e_2 \wedge e_3 \wedge
e_\infty $ is computed as 
\begin{alltt}
In[6]:= OuterProduct[e[0], e[1], e[2], e[3], e[\(\infty\)]
Out[6]:= -e[1,2,3] + e[0,1,2,3,\(\infty\)]
\end{alltt}
That is, $I_5 = -e_1 e_2 e_3 + e_0 e_1 e_2 e_3 e_\infty$.

The pseudoscalar is predefined in the package as \texttt{I5}:
\begin{alltt}
In[7]:= I5
Out[7]:= -e[1,2,3] + e[0,1,2,3,\(\infty\)]
\end{alltt}

\section{Representation of geometric objects}
\label{sec:usability}

Examples of direct and dual representations in the conformal space of
some geometric objects are presented in what follows.

\subsection{Points}
A point at the location
${\mathbf x} = (x, y, z) \in \mathbb{R}^3$ is represented by the
vector
\[
  x = e_0 + {\mathbf x} + \frac{1}{2} {\mathbf x}^2 e_\infty ,
\]
in $\mathbb{G}^{4,1}$.

We can immediately check that $x^2= xx = \| x \|^2 =0$:
\begin{alltt}
In[8]:= X = x1 e[1] + x2 x[2] + x3 e[3];
In[9]:= x = e[0] + X + (GeometricProduct[X,X]/2) e[\(\infty\)];
In[10]:= GeometricProduct[x, x] // Simplify
Out[10]:= 0
\end{alltt}

\subsection{Direct representation of lines}

A line $L$ passing through two points $p_1$ and $p_2$ is represented
by
\[
  L = p_1 \wedge p_2 \wedge e_\infty .
\]
Thus if $p$ is a point in the line, it satisfies
\begin{equation}
  \label{eqn:line}
  p \wedge L = 0 .
\end{equation}

Let ${\mathbf x} = (x, y, z)$, ${\mathbf x}_1 = (x_1, y_1, z_1)$,
${\mathbf x}_2 = (x_2, y_2, z_2)$,
$p = e_0 + {\mathbf x} + \frac{{\mathbf x}^2}{2} e_\infty$,
$p_1 = e_0 + {\mathbf x}_1 + \frac{{\mathbf x}_1 ^2}{2} e_\infty$ and
$p_2 = e_0 + {\mathbf x}_2 + \frac{{\mathbf x}_2 ^2}{2} e_\infty$. The
equation of the line passing through $p_1$ and $p_2$ can be obtained
as follows:
\begin{alltt}
In[11]:= X = x e[1] + y x[2] + z e[3];
In[12]:= Y = x1 e[1] + y1 x[2] + z1 e[3];
In[13]:= Z = x2 e[1] + y2 x[2] + z2 e[3];
In[14]:= p = e[0] + X + (Magnitude[X]^2/2) e[\(\infty\)];
In[15]:= p1 = e[0] + Y + (Magnitude[Y]^2/2) e[\(\infty\)];
In[16]:= p2 = e[0] + Z + (Magnitude[Z]^2/2) e[\(\infty\)];
In[17]:= line = OuterProduct[p, p1, p2, e[\(\infty\)]] // FullSimplify
Out[17]:= (-x1 y + x2 y + x y1 - x2 y1 - x y2 + x1 y2) e[1,2] + 
(-x1 z + x2 z + x z1 - x2 z1 - x z2 + x1 z2) e[1,3] + 
(-y1 z + y2 z + y z1 - y2 z1 - y z2 + y1 z2) e[2, 3]
(x2 y+x y1-x2 y1-x y2+x1(-y+y2)) e[0,1,2,\(\infty\)] +
(-x1 z+x2 z+x z1-x2 z1-x z2+x1 z2) e[0,1,3,\(\infty\)] + 
(-y1 z+y2 z+y z1-y2 z1-y z2+y1 z2) e[0,2,3,\(\infty\)] +
(-x2 y1 z+x1 y2 z+x2 y z1-x y2 z1-x1 y z2+x y1 z2) e[1,2,3,\(\infty\)]
\end{alltt}
Then from \eqref{eqn:line}, all the coefficients of the 2-vectors and
the 4-vectors must be set to zero:
\begin{alltt}
In[18]:= Solve[ \(\{\)Coefficient[line,e[1,2] == 0,
Coefficient[line,e[1,3] == 0,
Coefficient[line,e[2,3] == 0,
Coefficient[line,e[0,1,2,\(\infty\)] == 0,  
Coefficient[line,e[0,1,3,\(\infty\)] == 0, 
Coefficient[line,e[0,2,3,\(\infty\)] == 0, 
Coefficient[line,e[1,2,3,\(\infty\)]] == 0\(\}\),\(\{\)x, y, z\(\}\)]
Out[18]:= Solve::svars: Equations may not give solutions for all
"solve" variables.
\(\{\)\(\{\)y -> -((x (-y1 + y2))/(x1 - x2)) - (x2 y1 - x1 y2)/(x1 - x2), 
  z -> -((x (-z1 + z2))/(x1 - x2)) - (x2 z1 - x1 z2)/(x1 - x2)\(\}\)\(\}\)
\end{alltt}
That is:
\begin{eqnarray*}
  y &=& \frac{x ( y_1 - y_2)}{x_1 - x_2} - \frac{x_2 y_1 - x_1 y_2}{x_1 - x_2} ,\\
  z &=& \frac{x ( z_1 - z_2)}{x_1 - x_2} - \frac{x_2 z_1 - x_1 z_2}{x_1 - x_2} .
\end{eqnarray*}
If we define $x = (x_1 - x_2) t$, the parametric equations of the line are:
\begin{eqnarray*}
  x &=& \left( x_1 - x_2 \right) t , \\
  y &=& \left( y_1 - y_2 \right) t - \frac{x_2 y_1 - x_1 y_2}{x_1 -
        x_2}  ,\\
  z &=& \left( z_1 - z_2 \right) t - \frac{x_2 z_1 - x_1 z_2}{x_1 -
        x_2} .
\end{eqnarray*}

\subsection{Direct representation of planes}

A plane $P$ passing through three points $p_1$, $p_2$ and $p_2$ is
represented by
\begin{equation}
  \label{eqn:P}
  P = p_1 \wedge p_2 \wedge p_3 \wedge e_\infty .
\end{equation}
If $p$ is a point in the line, its equation if given as
\begin{equation}
  \label{eqn:plane}
  p \wedge P = 0 .
\end{equation}
As above, let ${\mathbf x} = (x, y, z)$,
${\mathbf x}_1 = (x_1, y_1, z_1)$, ${\mathbf x}_2 = (x_2, y_2, z_2)$,
${\mathbf x}_3 = (x_3, y_3, z_3)$, and
\begin{subequations}
  \label{eqn:p1p2p3}
  \begin{align}
    p &= e_0 + {\mathbf x} + \frac{{\mathbf x}^2}{2} e_\infty ,  \\
    p_1 &= e_0 + {\mathbf x}_1 + \frac{{\mathbf x}_1 ^2}{2} e_\infty , \\
    p_2 &= e_0 + {\mathbf x}_2 + \frac{{\mathbf x}_2 ^2}{2} e_\infty ,\\
    p_3 &= e_0 + {\mathbf x}_3 + \frac{{\mathbf x}_3 ^2}{2} e_\infty .
  \end{align}
\end{subequations}
In 3D, the equation of the plane passing through ${\mathbf x}_1$,
${\mathbf x}_2 $ and ${\mathbf x}_3$ can be obtained as follows:
\begin{alltt}
In[19]:= X = x e[1] + y x[2] + z e[3];
In[20]:= Y = x1 e[1] + y1 x[2] + z1 e[3];
In[21]:= Z = x2 e[1] + y2 x[2] + z2 e[3];
In[22]:= W = x3 e[1] + y3 x[2] + z3 e[3];
In[23]:= p = e[0] + X + (Magnitude[X]^2/2) e[\(\infty\)];
In[24]:= p1 = e[0] + Y + (Magnitude[Y]^2/2) e[\(\infty\)];
In[25]:= p2 = e[0] + Z + (Magnitude[Z]^2/2) e[\(\infty\)];
In[26]:= p3 = e[0] + W + (Magnitude[W]^2/2) e[\(\infty\)];
In[27]:= plane = OuterProduct[p, p1, p2, p3, e[\(\infty\)]] // FullSimplify
Out[27]:= -((-x1 y2 z + x1 y3 z + x y2 z1 - x y3 z1 + x1 y z2 - 
x y1 z2 + x y3 z2 - x1 y3 z2 + x3 (-y1 z + y2 z + y z1 - 
y2 z1 - y z2 + y1 z2) + (-x1 y + x y1 - x y2 + x1 y2) z3 + 
x2 (-y3 z - y z1 + y3 z1 + y1 (z - z3) + y z3))
(e[1, 2, 3] - e[0, 1, 2, 3, \(\infty\)]))
\end{alltt}
By equating the coefficient of $e_1  e_2  e_3 - e_0 e_1 e_2  e_3
e_\infty$ ($= I_5$) to
zero, and rearranging terms, the equation of the plane is:
\begin{eqnarray}
  \label{eqn:panedirect}
     \left( (z_3 - z_2) y_1 + (z_1 - z_3) y_2 + (z_2 - z_1) y_3 \right) x
     + && \nonumber \\
     \left( (z_2 - z_3) x_1 + (z_3 - z_1) x_2 + (z_1 - z_2) x_3 \right) y +
       && \nonumber \\
     \left( (y_3 - y_2) x_1 + (y_1 - y_3) x_2 + (y_2 - y_1) x_3 \right) z +
       && \nonumber \\
     (x_2 y_3 -  x_3 y_2 ) z_1 + (x_3 y_1 - x_1 y_3) z_2+( x_1 y_2 -
     x_2 y_1) z_3 = 0 .
\end{eqnarray}

\subsection{Direct representation of spheres}

A sphere passing through four points $p_1$, $p_2$, $p_3$ and $p_4$, is
represented by
\[
S = p_1 \wedge p_2 \wedge p_3 \wedge p_4 ,
\]
and its equations is
\[
p \wedge S = 0 .
\]

Consider ${\mathbf x}$, ${\mathbf x}_1$, ${\mathbf x}_2$,
${\mathbf x}_3$, $p$, $p_1$, $p_2$ and $p_3$ from the above examples
and ${\mathbf x}_4 = (x_4, y_4, z_4)$,
$p_4 = e_0 + {\mathbf x}_4 + \frac{{\mathbf x}_4 ^2}{2}
e_\infty$. 
\begin{alltt}
In[28]:= V = x4 e[1] + y4 x[2] + z4 e[3];
In[29]:= p4 = e[0] + V + (Magnitude[V]^2/2) e[\(\infty\)];
In[30]:= sphere = OuterProduct[p, p1, p2, p3, p4] 
Out[30]:= (.....) (e[1, 2, 3] - e[0, 1, 2, 3, \(\infty\)])
\end{alltt}

The result produces a large output (omitted), times
$e_1 e_2 e_3 - e_0 e_1 e_2 e_3 e_\infty = I_5$. As an example, consider the
sphere passing through the points ${\mathbf x}_1 = (1, -1, 3)$,
${\mathbf x}_2 = (4, 1, -2)$, ${\mathbf x}_3 = (-1, -1, 1)$,
${\mathbf x}_4 = (1, 1, 1)$:

\begin{figure}[h!]
  \centerline{\includegraphics[width=7cm]{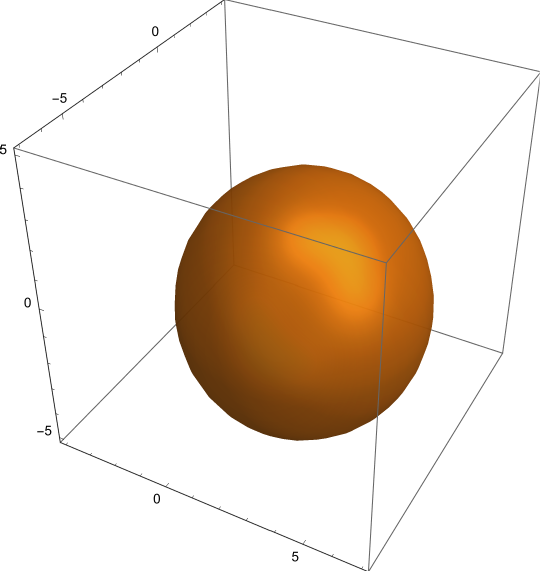}}
\caption{\label{fig:sphere}}
\end{figure}

\begin{alltt}
In[31]:= sphere = sphere /. \(\{\)x1-> 1, y1-> -1, z1-> 3, x2-> 4, y2-> 1, 
z2-> -2, x3-> -1, y3-> -1, z3-> 1, x4-> 1, y4-> 1, z4-> 1\(\}\)
Out[31]:= -12 (-4+(-5+x) x+y (5+y)+z+ z^2) (e[1, 2, 3] - e[0, 1, 2, 3, \(\infty\)])
In[32] := sphere = Coefficient[sphere, (e[1, 2, 3] - e[0, 1, 2, 3, \(\infty\)])]]
Out[32] := -12 (-4+(-5+x) x+y (5+y)+z+ z^2)
\end{alltt}
The equation of the sphere is then:
\[
(x-5) x + (y+5) y + (z+1) z -4 = 0.
\]
And
\begin{alltt}
In[33]:= ContourPlot3D[sphere == 0,\(\{\)-4,7\(\}\),\(\{\)-7,4\(\}\),\(\{\)-5,5\(\}\), Mesh->None,
         ContourStyle -> Directive[Opacity[0.8], Specularity[White, 30]]]
\end{alltt}
produces the sphere displayed in Fig.\ref{fig:sphere}.

\subsection{Dual representation of planes}

Consider a plane passing through the points $p_1$, $p_2$ and $p_3$
given in Equations \ref{eqn:p1p2p3}(b)-(d). This plane is represented
by the blade \eqref{eqn:P}:
\begin{alltt}
In[34]:= P = OuterProduct[p1, p2, p3, e[\(\infty\)]] // GFactor
Out[34]:= (-x2 y1 + x3 y1 + x1 y2 - x3 y2 - x1 y3 + x2 y3) e[1, 2]
+ (-x2 z1 + x3 z1 + x1 z2 - x3 z2 - x1 z3 + x2 z3) e[1, 3]
+ (-y2 z1 + y3 z1 + y1 z2 - y3 z2 - y1 z3 + y2 z3) e[2, 3]
+ (-x2 y1+x3 y1+x1 y2-x3 y2-x1 y3+x2 y3) e[0,1,2,\(\infty\)]
+ (-x2 z1+x3 z1+x1 z2-x3 z2-x1 z3+x2 z3) e[0,1,3,\(\infty\)]
+ (-y2 z1+y3 z1+y1 z2-y3 z2-y1 z3+y2 z3) e[0,2,3,\(\infty\)]
+ (-x3 y2 z1+x2 y3 z1+x3 y1 z2-x1 y3 z2-x2 y1 z3+x1 y2 z3) e[1,2,3,\(\infty\)]
\end{alltt}
Where \texttt{GFactor[]} is a function defined in the package that
groups terms with common \texttt{e[..]}'s.

The dual $P^\ast$ can be calculated using the \texttt{Dual[]} function
defined in the package, or:
\begin{alltt}
In[35]:= Pdual = -InnerProduct[P,I5]
Out[34]:= (y2 z1-y3 z1-y1 z2+y3 z2+y1 z3-y2 z3) e[1]
+ (-x2 z1+x3 z1+x1 z2-x3 z2-x1 z3+x2 z3) e[2]
+ (x2 y1-x3 y1-x1 y2+x3 y2+x1 y3-x2 y3) e[3]
+ (x3 y2 z1-x2 y3 z1-x3 y1 z2+x1 y3 z2+x2 y1 z3-x1 y2 z3) e[\(\infty\)]
\end{alltt}
By defining $h$ and ${\mathbf n}=(n_1, n_2, n_3)$, where
\begin{eqnarray*}
  n_1 &=& y_2 z_1 - y_3 z_1 - y_1 z_2 + y_3 z_2 + y_1 z_3 - y_2 z_3,
  \\
      &=& (z_3 - z_2) y_1  + (z_1 - z_3) y_2 + (z_2 - z_1) y_3 \\
  n_2 &=& x_3 z_1 - x_2 z_1 + x_1 z_2 - x_3 z_2 - x_1 z_3 + x_2 z_3,
  \\
      &=& (z_2 - z_3) x_1 + (z_3 - z_1) x_2 + (z_1 - z_2) x_3 \\
  n_3 &=& x_2 y_1 - x_3 y_1 - x_1 y_2 + x_3 y_2 + x_1 y_3 - x_2 y_3,
  \\
      &=& (y_3 - y_2) x_1 + (y_1 - y_3) x_2 + (y_2 - y_1) x_3 \\
 h &=&  (x_3 y_2 z_1 - x_2 y_3 z_1 - x_3 y_1 z_2 + x_1 y_3 z_2 + x_2
        y_1 z_3 - x_1 y_2 z_3), \\
      &=& ( x_3 y_2  - x_2 y_3) z_1 +  ( x_1 y_3  - x_3 y_1) z_2 +
          (x_2 y_1  - x_1 y_2) z_3 ,
\end{eqnarray*}
then
\[
 P^\ast = n_1 e_1 + n_2 e_2 + n_3 e_3 - h e_\infty = {\mathbf n} - h
 e_\infty .
\]
If ${\mathbf x} = (x, y, z)$ is a point in the plane, represented by
$p$ in (\ref{eqn:p1p2p3}a), the equation of
the plane is
\[
p \cdot P^\ast = 0,
\]
\begin{alltt}
In[36]:= pdual = n1 e[1] + n2 e[2] + n3 e[3] - h e[\(\infty\)];
In[37]:= InnerProduct[p,pdual]
Out[37]:= -h + n1 x + n2 y + n3 z
\end{alltt}
that is
\[
 n_1 x + n_2 y + n_3 z - h = 0,
\]
which, as expected, equals \eqref{eqn:panedirect}.

\subsection{Dual representation of spheres}

Consider the point $p_1$ given in (\ref{eqn:p1p2p3}b); a sphere of
radius $r$ centered at $p_1$ is represented as
\begin{equation}
  \label{eqn:sph}
  S = p_1 - \frac{r^2}{2} e_\infty .
\end{equation}
If ${\mathbf x} = (x, y, z)$ is a point in the sphere, represented by
$p$ in (\ref{eqn:p1p2p3}a), we notice that
$p \cdot S$ is
\begin{alltt}
In[38]:= S = p1 -  r^2/2 e[\(\infty\)];
In[39]:= InnerProduct[p,S] // FullSimplify
Out[39]:= 1/2 (r^2 - (x-x1)^2 - (y-y1)^2 - (z-z1)^2)
\end{alltt}
That is
$p \cdot S = \left( r^2 - \| {\mathbf x} - {\mathbf x}_1\| ^2 \right)
/ 2$, and $p \cdot S = 0$:
\[
\| {\mathbf x} - {\mathbf x}_1\|^2  = r^2 ,
\]
is the equation of the sphere with center $ {\mathbf x}_1$ and radius
$r$, and $S$ in \eqref{eqn:sph} is the dual representation of this
sphere.

\section{Transformations}

One of the advantages of the conformal model is that conformal
tranformations in $\mathbb{R}^3$ can be represented by orthogonal
transformations in $\mathbb{G}^{4,1}$. In what follows
\texttt{CGAlgebra} is applied to describe translations, rotations and
rigid motions.

\subsection{Translations}

Translations are obtained as reflections by two parallel planes. The
translation of a point ${\mathbf p}$ in $\mathbb{R}^3$ by the vector
${\mathbf t} = t_1 e_1 + t_2 e_3 +t_3 e_3$ is
\[
  T_{\mathbf t} \; {\mathbf p} \; T_{\mathbf t} ^{-1} ,
\]
where
\[
  T_{\mathbf t} = 1- \frac{1}{2} {\mathbf t} e_\infty.
\]
We can easily check that  $T_{\mathbf t} ^{-1} = T_{-{\mathbf t}}$:
\begin{alltt}
In[40]:= t = t1 e[1] + t2 e[2] + t3 e[3];
In[41]:= Tt = 1-GeometricProduct[t,e[\(\infty\)]]/2
Out[41]:= 1 - (t1 e[1,\(\infty\)]]- t2 e[2,\(\infty\)]- t3 e[3,\(\infty\)])/2
In[42]:= Tti = MultivectorInverse[Tt]
Out[42]:= 1+ (t1 e[1,\(\infty\)]]+ t2 e[2,\(\infty\)]+ t3 e[3,\(\infty\)])/2
In[43]:= Tti == 1-1/2 GeometricProduct[-t,e[\(\infty\)]]  // Simplify
Out[43]:= True
\end{alltt}
where the function \texttt{MultivectorInverse[]} defined in the
package returns the Inverse (if it exists) of a multivector. Some
simple examples are as follows.

\begin{itemize}
\item Translation of the origin $e_0$ by the vector ${\mathbf t}$:
\begin{alltt}
In[44]:= GeometricProduct[Tt,e[0],Tti];
Out[44]:= e[0]+t1 e[1]+t2 e[2]+t3 e[3]+1/2(t1^2+t2^2+t3^2) e[\(\infty\)]
\end{alltt}
thus $T_{\mathbf t} e_0 T_{\mathbf t} ^{-1} = e_0 + {\mathbf t} +
\frac{1}{2} {\mathbf t} ^2 e_{\infty} = t$, where $t$ is the
representation of ${\mathbf t}$ in $\mathbb{G}^{4,1}$.

\item Translation of the infinity $e_\infty$:
\begin{alltt}
In[45]:= GeometricProduct[Tt,e[\(\infty\)],Tti];
Out[45]:= e[\(\infty\)]
\end{alltt}

\item Translation of a vector ${\mathbf x} = (x_1, x_2, x_3)$:
\begin{alltt}
In[46]:= x= x1 e[1] + x2 e[2] + x3 e[3];
In[47]:= GeometricProduct[Tt,x,Tti];
Out[47]:= x1 e[1]+x2 e[2]+x3 e[3]+(t1 x1+t2 x2+t3 x3) e[\(\infty\)]
\end{alltt}
  that is,
  $T_{\mathbf t} {\mathbf x} T_{\mathbf t} ^{-1} = {\mathbf x} +
  \left( {\mathbf x} \cdot {\mathbf t} \right) e_\infty$.
  
\end{itemize}

\subsection{Rotations}

The rotation of a vector ${\mathbf x} \in \mathbb{R}^3$ around an axis
orthogonal to the vectors ${\mathbf a}$ and ${\mathbf b}$ is given by
\begin{equation}
   \label{eqn:rrp}
  {\mathbf x}^\prime = R \; {\mathbf x} \; R^{-1},
\end{equation}
where $R = {\mathbf a} {\mathbf b}$.

The rotation plane is specified by the vectors ${\mathbf a}$ and
${\mathbf b}$ and if a rotation angle $\phi$ is given, then
\[
  {\mathbf x}^\prime = \left( \cos \left( \phi / 2 \right) -
    \hat{A} \sin \left( \phi / 2 \right)\right) {\mathbf x}
   \left( \cos \left( \phi / 2 \right) + \hat{A} \sin \left( \phi /
      2 \right) \right) ,
\]
where
\[
  \hat{A} = \frac{{\mathbf a} \wedge {\mathbf b}}{| {\mathbf a} \wedge
    {\mathbf b} |} .
\]
In (\ref{eqn:rrp}) the rotation angle is the angle between ${\mathbf a}$ and
${\mathbf b}$.

We can easily verify that $e_0$ and $e_\infty$ are not affected by a
rotation:
\begin{alltt}
In[48]:= a = a1 e[1] + a2 e[2] + a3 e[3];
In[49]:= b = b1 e[1] + b2 e[2] + b3 e[3];
In[50]:= R = GeometricProduct[a,b]; 
In[51]:= GeometricProduct[R,e[0],MultivectorInverse[R]];
Out[51]:= e[0]
In[52]:= GeometricProduct[R,e[\(\infty\)],MultivectorInverse[R]];
Out[52]:= e[\(\infty\)]
\end{alltt}

\section{Additional predefined functions}

The basic functions listed in Table \ref{tbl:t1} are the starting
point to define more functions depending on the applications. To
make thing easier, some complementary useful functions are defined in
\texttt{CGAlgebra}, and they are listed in Table \ref{tbl:t2}.

\begin{table}[ht!]
  \caption{\label{tbl:t2}Complementary functions of \texttt{CGAlgebra}}
\begin{center}
\begin{tabular}{|l|l|}
  \hline 
  \textbf{Expression} & \textbf{Output} \tabularnewline
               \hline 
               \hline 
               \texttt{Magnitude[A]} & $A^2$ \\
  \hline 
  \texttt{Reversion[A]} & The reversion of the multivector $A$\\
  \hline
  \texttt{Involution[A]} & The grade involution ($\dagger$) of the multivector $A$.
  \\
  \hline 
  \texttt{MultivectorInverse[A]} & The
                                                    inverse (if it
                                                    exists) of the multivector $A$.\\
  \hline 
  \texttt{GradeQ[A,k]} & \texttt{True} if $A$ is a $k$-vector.\\
  \hline
  \texttt{Dual[A]} & The dual of the multivector $A$\\
  \hline
  \multirow{5}{*}{\texttt{Rotation[x,a,b,\(\theta\)]}} & Rotation of the vector
                                                    $x$ by an angle
                                                    $\theta$ along\\
  ~ & the plane defined by the vectors $a$ and $b$. \\
  ~ & The sense of the rotation is from $a$ to $b$. \\
  ~ & The default value of $\theta$ is the angle between\\
  ~ & $a$ and $b$.\\
  \hline
\multirow{3}{*}{\texttt{ToVector[v]}} & The $\mathbb{R}^3$
                                        vector \texttt{v=x e[1]+y e[2]+z e[3]} \\
 ~ & transformed to the \emph{Mathematica} \\
  ~ & input form \texttt{\(\{\)x,y,z\(\}\)}. \\
  \hline
\multirow{2}{*}{\texttt{GFactor[A]}} & Factors terms of the expression $A$\\
 ~ & with common $e[i,j,k]$'s. \\
  \hline
\texttt{I5} & The pseudoscalar $I$ of $\mathbb{G}^{4,1}$\\
  \hline
\texttt{I5i} & $I^{-1}$\\
 \hline
\end{tabular}
\end{center}
\end{table}

With the basic and complementary functions described in Tables
\ref{tbl:t1} and \ref{tbl:t2}, any other required function can be
built to enlarge the package using the capability of the
\emph{Wolfram} language to define our own functions. Considere for
example the \emph{sphere inversion}.

Let $S$ be an sphere with center $p$ and radius $r$. From
(\ref{eqn:sph}) we have:
\[
  S = p - \frac{r^2}{2} e_\infty .
\]
The inversion of a point $x$ in $\mathbb{G}^{4,1}$, with respect to
the sphere $S$ is
\[
-S x S^{-1} .
\]
The \emph{Mathematica} expression that represents the inversor
operator is defined as follows:
\begin{alltt}
In[53]:= Clear[x,p,r]
In[54]:= Inversor[x_,p_,r_]:= 
 -GeometricProduct[p-r^2 e[\(\infty\)]/2,x,MultivectorInverse[p-r^2 e[\(\infty\)]/2]]
\end{alltt}
Thus, the user-defined function \texttt{Inversor[x,p,r]} inverts a
point $x$ with respect to a sphere centered at $p$ and with radius
$r$. Let us try \texttt{Inversor[]} by considering inversions with
respect to a sphere centered at the origin, that is $p=e[0]$.

The inversion of the origin $e_0$ is $e_\infty$:
\begin{alltt}
In[55]:= Inversor[e[0],e[0],r]
Out[55]:= r^2 e[\(\infty\)]/2
\end{alltt}
The infinity $e_\infty$ is inverted to the origin:
\begin{alltt}
In[56]:= Inversor[e[\(\infty\)],e[0],r]
Out[56]:= 2 e[0]/r^2
\end{alltt}
Vectors in $\mathbb{R}^3$ are unaffected by the inversion:
\begin{alltt}
In[57]:= v = v1 e[1] + v2 e[2] + v3 e[3]; 
In[58]:= Inversor[v,e[0],r]
Out[58]:= v1 e[1] + v2 e[2] + v3 e[3]
\end{alltt}

In this way many other expressions can be defined. Please feel free to
contact the authors if further assistance is required to implement
\text{CGAlgebra} to a particular application.

\end{document}